\documentclass{osa-article}

\journal{osac}

\articletype{Research Article}

\usepackage{dcolumn}
\usepackage{graphicx}% Include figure files
\usepackage{dcolumn}% Align table columns on decimal point
\usepackage{bm}% bold math
\usepackage{physics}
\usepackage{color}
\usepackage{ulem}

\usepackage{graphicx}% Include figure files
\usepackage{bm}% bold math
\usepackage{physics}

\begin{document}

\title{Quantum Coherence of Orbital Angular Momentum Multiplexed Continuous-Variable Entangled State}

\author{Hong Wen,\authormark{1,2} Li Zeng,\authormark{1,2} Rong Ma,\authormark{1,2} Haijun Kang,\authormark{1,2} Jun Liu,\authormark{1,2} Zhongzhong Qin,\authormark{1,2,*} and Xiaolong Su\authormark{1,2}}

\address{\authormark{1}State Key Laboratory of Quantum Optics and Quantum Optics Devices, Institute of Opto-Electronics, Shanxi University, Taiyuan, 030006, People's Republic of China\\
\authormark{2}Collaborative Innovation Center of Extreme Optics, Shanxi University, Taiyuan,Shanxi 030006, People's Republic of China}

\email{\authormark{*}zzqin@sxu.edu.cn} 

\begin{abstract}
Orbital angular momentum (OAM) multiplexed entangled state is an important quantum resource for high dimensional quantum information processing. In this paper, we experimentally quantify quantum coherence of OAM multiplexed continuous-variable (CV) entangled state and characterize its evolution in a noisy environment. We show that the quantum coherence of the OAM multiplexed CV entangled state carrying topological charges $l=1$ and $l=2$ are the same as that of the Gaussian mode with $l=0$ in a noisy channel. Furthermore, we show that the quantum coherence of OAM multiplexed entangled state is robust to noise, even though the sudden death of entanglement is observed. Our results provide reference for applying quantum coherence of OAM multiplexed CV entangled state in a noisy environment.
\end{abstract}

%%%%%%%%%%%%%%%%%%%%%%%%%%  body  %%%%%%%%%%%%%%%%%%%%%%%%%%
\section{Introduction}
Quantum coherence, which characterizes the quantumness and underpins quantum correlation in quantum systems, plays a significant role in quantum information processing \cite{Baumgratz2014}. Recently, quantum coherence has been identified as important quantum resource besides quantum entanglement and steering, and has attracted rapidly increasing interests \cite{Streltsov2017,Chitambar2019}. In 2014, Baumgratz \textit{et al.} established a framework to quantify quantum coherence by referring to the method of quantifying entanglement \cite{Baumgratz2014}. The quantum coherence of a quantum state is defined as the minimum distance between the quantum state and an incoherent state in the Hilbert space, and can be quantified by relative entropy and $l_{1}$-norm \cite{Baumgratz2014}. Besides, it has been shown that quantum coherence can also be quantified by Fisher information \cite{Feng2017}, skew information entropy \cite{Yu2017}, Tsallis relative $\alpha$ entropy \cite{Rastegin2016}, robustness \cite{Napoli2016}, and so on. Furthermore, quantum coherence with infinite-dimensional systems, i.e. continuous-variable (CV) quantum states, can be quantified by relative entropy \cite{Fan2016}. Both theoretical investigations and experimental demonstrations of quantum coherence have achieved significant progresses \cite{Xu2016, Buono2016, Albarelli2017,Yuan2017D,Wu2017,Gao2018,Wu2018,Zhang2019,Xu2020,ZhangPRJ2021}. Recently, we have experimentally demonstrated the robustness of Gaussian quantum coherence in quantum channels \cite{KangPRJ2021}, as well as the conversion of local and correlated Gaussian quantum coherence \cite{KangOL2021}. 

As an important quantum resource, Einstein-Podolsky-Rosen (EPR) entangled state has been widely applied in quantum communication, quantum computation, and quantum precision measurement \cite{EPREntanglement,BraunsteinRMP,KimbleQuanInt,WeedbrookRMP,PhysicsReport,HaoCPB2021,SuSciCH2021}. Besides optical parametric amplifier, four-wave mixing (FWM) process in warm alkali vapor cell is another efficient method to generate CV EPR entangled state \cite{EntangledImages,MaOL2018}. The FWM process has been widely used in quantum state engineering \cite{QinLight,QinPRL,QinOL2014}, quantum beam splitter \cite{LiuOL2019}, and quantum precision measurement \cite{SU11,PooserOptica}. Especially, spatial-multi-mode advantage of the FWM process, attributed to its cavity-free configuration, makes it an ideal method to generate entangled images \cite{EntangledImages}. Orbital angular momentum (OAM) multiplexing of light has been found to be an efficient way to improve data-carrying capacity in both classical and quantum communications due to its infinite range of possibly achievable topological charges \cite{AllenOAM,OAMMultiplexing}. Recently, OAM multiplexed bipartite and multipartite CV entangled states have been generated based on the FWM process \cite{JingBiOAM,JingTriOAM,JingHexaOAM}, and they have been applied in OAM multiplexed quantum teleportation \cite{JingQuanTele} and quantum dense coding \cite{JingQDC}. It is essential to distribute OAM multiplexed quantum resources in quantum channels to realize these quantum information processing protocols. Recently, our group has experimentally demonstrated that quantum entanglement and quantum steering of the OAM multiplexed states carrying topological charges $l=1$ and $l=2$ are the same as that of the Gaussian mode with $l=0$ in lossy and noisy channels \cite{PRJ2022}. However, it remains unclear whether the quantum coherence of OAM multiplexed CV entangled state is also as robust as that of the Gaussian mode with $l=0$.

Here, we experimentally quantify quantum coherence of OAM multiplexed CV entangled state and characterize its evolution in a noisy channel. We show that quantum coherence of CV entangled state carrying topological charges $l=1$ and $l=2$ are as robust against loss and noise as that of Gaussian mode entangled state with $l=0$. More interestingly, the quantum coherence of CV entangled state always exists unless one mode is completely lost, while sudden death of entanglement is observed in the presence of certain amounts of loss and noise. Our results pave the way for applying the quantum coherence of OAM multiplexed entangled state in high data-carrying capacity quantum communication protocols.

\section{Theory}
The Hamiltonian of the OAM multiplexed FWM process can be expressed as \cite{PRJ2022}:
\begin{equation}
\hat{H}=\sum_{l}i\hbar\gamma_{l}\hat{a}^{\dagger}_{l,P}\hat{a}^{\dagger}_{-l,C}+h.c.
\end{equation}
where \(\gamma_{l}\) is defined as the interaction strength of each OAM pair. \(\hat{a}^{\dagger}_{l,P}\) and \(\hat{a}^{\dagger}_{-l,C}\) are the creation operators related to OAM modes of the Pr and Conj fields, respectively. Since the pump field does not carry OAM (\(l=0\)), the topological charges of the Pr and Conj fields are opposite. The output state of the OAM multiplexed FWM process is:
\begin{equation}
\ket{\Psi}_{out}=\ket{\Psi}_{-l}\otimes\cdots\otimes\ket{\Psi}_{0}\otimes\cdots\otimes\ket{\Psi}_{l}
\end{equation}
where \(\ket{\Psi}_{l}\) presents a series of independent OAM multiplexed CV EPR entangled states of $\ket{\psi_{l,P}}$ and $\ket{\psi_{-l,C}}$ generated in the FWM process. $\ket{\psi_{l,P}}$ and $\ket{\psi_{-l,C}}$ represent Pr field carrying topological charge $l$ and Conj field carrying topological charge $-l$ , respectively.

	A Gaussian state  $\hat{\rho}(\bar{\mathbf{x}},\mathbf{V})$ can be completely represented by the displacement $\bar{\mathbf{x}}$ and the covariance matrix $\mathbf{V}$ in phase space, which correspond to the first and second statistical moments of the quadrature operators, respectively \cite{WeedbrookRMP,PhysicsReport}. The displacement $\mathbf{\bar{x}}=\langle 
	\hat{x}\rangle $, where	
	$\hat{x}\equiv (\hat{X}_{-l,C}, \hat{Y}_{-l,C}, \hat{X}_{l,P}, \hat{Y}_{l,P})^{T}$, $\hat{X}=\hat{a}+\hat{a}^{\dag}$ and $\hat{Y}=(\hat{a}-\hat{a}^{\dag})/i$ are the amplitude and phase quadratures of an optical mode, respectively, and $T$ denotes transpose.
	The elements of covariance matrix $\mathbf{V}$ are defined as
	$\mathbf{V}_{ij}=\frac{1}{2}\langle \hat{x}_{i}\hat{x}_{j}+\hat{x}_{j}\hat{x}_{i}\rangle -\langle \hat{x}_{i}\rangle \langle \hat{x}_{j}\rangle$. The Gaussian quantum coherence of EPR entangled state is expressed by \cite{Xu2016}
	\begin{equation}
	\mathcal{C}_{rel.~ent.}\left[\hat{\rho}(\bar{\mathbf{x}},\mathbf{V})\right] =S\left[\hat{\rho}(\bar{\mathbf{x}}_{th},\mathbf{V}
	_{th})\right] -S\left[\hat{\rho}(\bar{\mathbf{x}},\mathbf{V}) \right],
	\end{equation}
	where $S\left[\hat{\rho}(\bar{\mathbf{x}},\mathbf{V})\right]\!=-\!\underset{i=1}{\overset{2}{\sum }}\!\left[\!\left(
	\frac{\nu_{i}-1}{2}\right)\!\log\!_{2}\!\left( \frac{\nu_{i}-1}{2}
	\right)\!-\!\left( \frac{\nu_{i}+1}{2}\right)\!\log\!_{2}\!\left( \frac{\nu
		_{i}+1}{2}\right)\!\right]$ and
	
	$S\left[ \hat{\rho}(\bar{\mathbf{x}}_{th},\mathbf{V} _{th})\right]\!\!=\!-\!\underset{i=1}{\overset{2}{\sum }}\!\left[\!\left(
	\frac{\mu_{i}-1}{2}\right)\!\log\!_{2}\!\left( \frac{\mu_{i}-1}{2}
	\right)\!-\!\left( \frac{\mu_{i}+1}{2}\right)\!\log\!_{2}\!\left( \frac{\mu_{i}+1}{2}\right)\!\right]$ are the von Neumann entropy of $\hat{\rho}(\bar{\mathbf{x}},\mathbf{V})$ and a thermal state $\hat{\rho}(\bar{\mathbf{x}}_{th},\mathbf{V}
	_{th})$, respectively.  
	$\nu_{i}$ and $\mu_{i}$ are the symplectic eigenvalues of $\mathbf{V}$
	and $\mathbf{V}_{th}$, respectively. The elements of the diagonal covariance matrix $\mathbf{V}_{th}$ are given by $\mathbf{V}$ with $V_{th}\ _{2i-1,2i-1}=V_{th}\ _{2i,2i}=\frac{1}{2}\left(V_{2i-1,2i-1}+V_{2i,2i}+\left[ \mathbf{\bar{x}}_{2i-1}\right] ^{2}+\left[\mathbf{\bar{x}}_{2i}\right] ^{2}\right)$.

\begin{figure}[h!]
\centering\includegraphics[width=10cm]{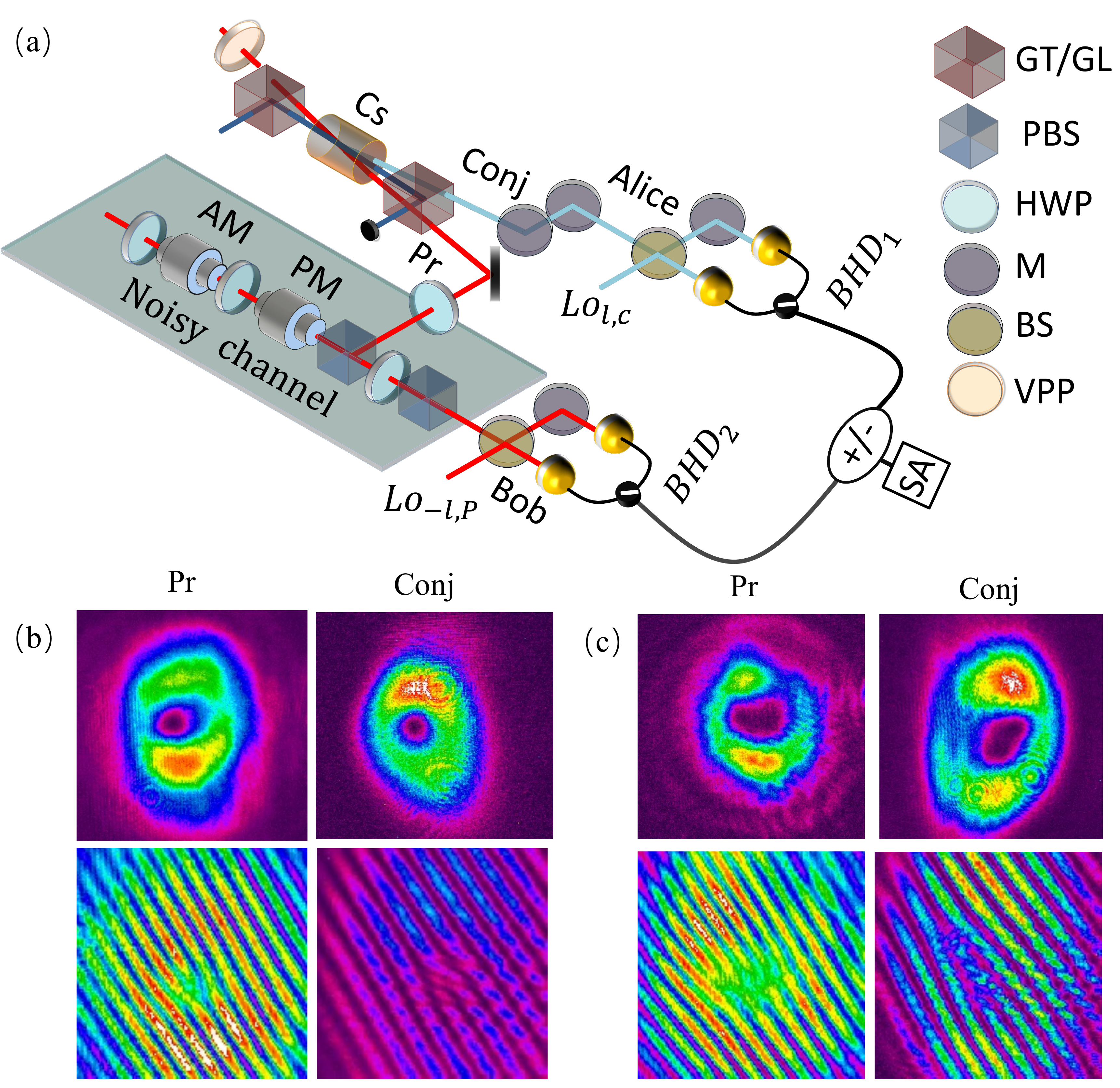}
\caption{(a) Experimental setup for generating and distributing quantum coherence of OAM multiplexed CV quantum entangled state in a noisy channel. Cs: cesium vapor cell; Pr: probe field; Conj: conjugate field; LO\(_{-l, P}\) and LO\(_{l, C}\): local oscillators of Pr and Conj fields; AM: amplitude modulator; PM: phase modulator; GL: Glan-laser polarizer; GT: Glan-Thompson polarizer; PBS: polarization beam splitter; HWP: half-wave plate; VPP: vortex phase plate; M: mirror; BS: 50:50 beam splitter; BHD$_{1}$, BHD$_{2}$: balanced homodyne detectors; SA: spectrum analyzer. (b) Beam patterns and interference patterns of the Pr and Conj fields for \(l=1\). (c) Beam patterns and interference patterns of the Pr and Conj fields for \(l=2\).}
\end{figure}
	
	In our experiment, the displacements $\bar{\mathbf{x}}$ of the EPR entangled state are zero, so the state can be completely represented by its covariance matrix $\mathbf{V}$. The covariance matrix of the OAM multiplexed entangled state after distribution in a noisy channel is 
	\begin{align}
	\mathbf{V}&=\left(
	\begin{array}{cccc}
	\mathbf{A} & \mathbf{C} \\
	\mathbf{C}^{T} & \mathbf{B}
	\end{array}%
	\right),
	\end{align}
	where $\mathbf{A}=\frac{V+V^{\prime}}{2}~\mathbf{I}$, $\mathbf{B}=[\eta\frac{V+V^{\prime}}{2}+(1-\eta)(1+\delta)]~\mathbf{I}$, $\mathbf{C}=\sqrt{\eta} \frac{V^{\prime}-V}{2}~\mathbf{Z}$, 
	$\mathbf{I}=\begin{pmatrix}	
	\begin{smallmatrix}	
	1 & 0 \\
	0 & 1	
	\end{smallmatrix}	
	\end{pmatrix}$ and
	$\mathbf{Z}=\begin{pmatrix}	
	\begin{smallmatrix}	
	1 & 0 \\
	0 & -1	
	\end{smallmatrix}	
	\end{pmatrix}$. 
	The submatrices $\mathbf{A}$ and $\mathbf{B}$ correspond to the states of Alice's and Bob's subsystems, respectively. $V$ and $V^{\prime}$ represent the variances of correlation and anti-correlation quadratures of the EPR entangled state, respectively. Note that $VV^{\prime} \geq 1$ is always satisfied according to the uncertainty principle, and the equality holds only for pure state. $\eta$ and $\delta$ (in the units of SNL) represent transmission efficiency and excess noise of the noisy channel, respectively. Therefore, $\delta=0$ represents a lossy but noiseless channel, while $\delta>0$ represents a noisy channel. 

The Peres-Horodecki criterion of positivity under partial transpose (PPT) criterion is a sufficient and necessary criterion to characterize the entanglement of CV bipartite entanglement \cite{PPTCriterion}. The PPT value is determined by  $\sqrt{\frac{\Gamma - \sqrt{\Gamma^2-4\textup{Det}\mathbf{V}}}{2}}$ for a bipartite Gaussian state, where $\Gamma=\textup{Det}\mathbf{A}+\textup{Det}\mathbf{B}-2\textup{Det}\mathbf{C}$. If the PPT value is smaller than 1, bipartite entanglement exists. Otherwise, it's a separable state. Furthermore, smaller PPT values represent stronger entanglement.

\begin{figure}[htbp]
	\centering
	\includegraphics[width=\linewidth]{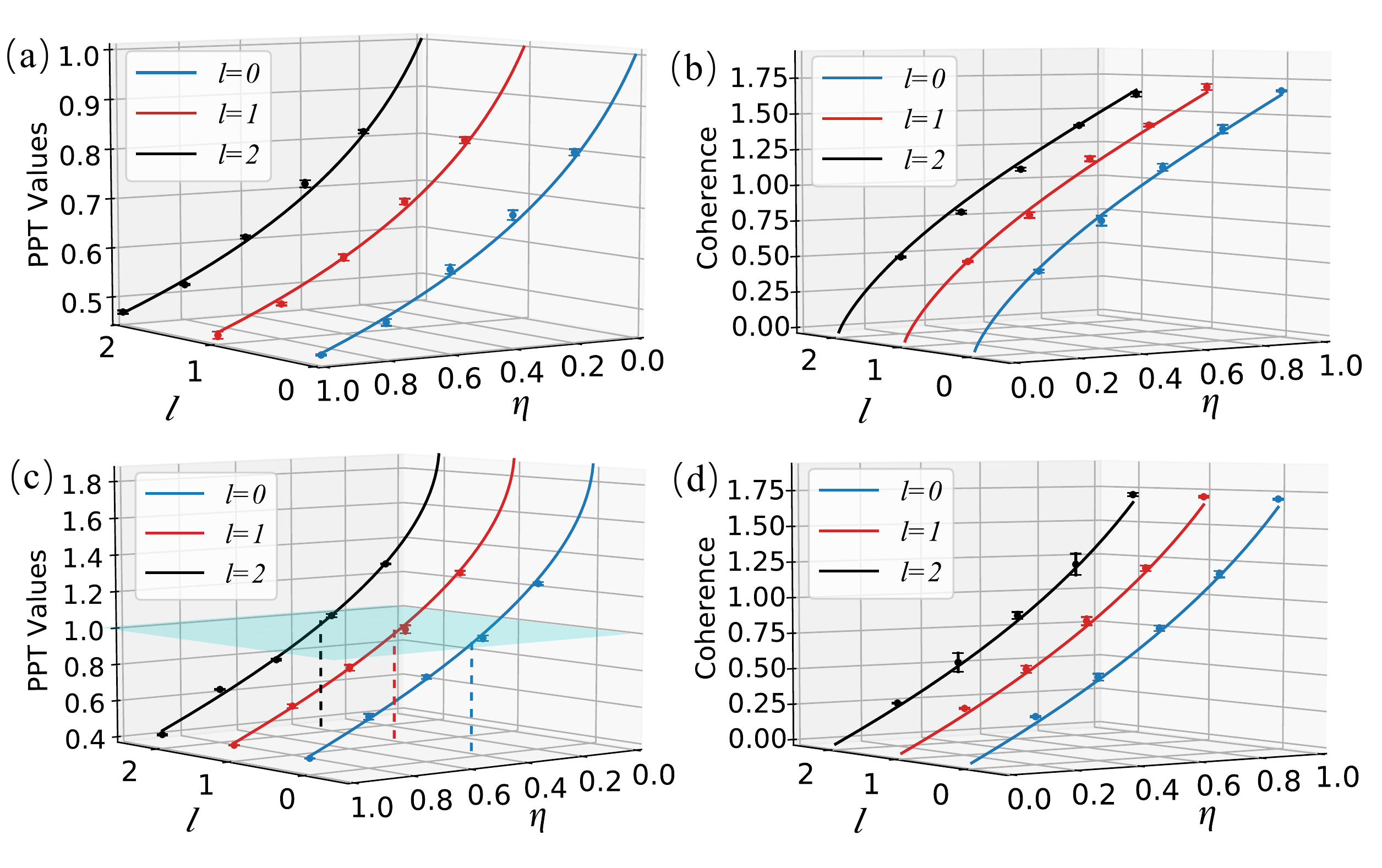}
	\caption{Dependence of PPT values and quantum coherence of the OAM multiplexed CV entangled states on the transmission efficiency $\eta$ for \(l=0\), \(l=1\) and \(l=2\). (a,b) Case for a lossy channel with the excess noise $\delta=0$ ; (c,d) Case for a noisy channel with the excess noise $\delta=1$. The initial PPT value is 0.46$\pm$0.01 at $\eta=1$. Curves and data points show theoretical predictions and experimental results, respectively. Error bars of experimental data represent one standard deviation and are obtained based on the statistics of the measured data. The light blue plane in (c) shows the boundary for entanglement where PPT value is equal to 1. The three vertical dashed lines indicate corresponding transmission efficiencies where sudden death of entanglement starts to appear.}
	\label{fig2}
\end{figure}

\section{Experimental setup}
Figure 1(a) shows the schematic of experimental setup. The Gaussian mode pump field and probe field carrying topological charge $l$ of OAM mode cross each other in the center of the cesium vapor cell at an angle of 6 mrad \cite{MaOL2018}. In this way, a conjugate field carrying topological charge $-l$ of OAM mode is generated on the other side of the pump field. The topological charge of OAM mode $l=1$ or $l=2$ is added to the probe field by passing it through a vortex phase plate. The pump field is filtered out by using a Glan-Thompson polarizer after the vapor cell. The Conj field is kept by Alice, while the Pr field is distributed to a remote quantum node owned by Bob through a noisy channel. The noisy channel is realized by overlapping the Pr field with an auxiliary field at a polarization beam splitter (PBS) followed by a half-wave plate (HWP) and a PBS. The auxiliary field, which is modulated by an amplitude modulator and a phase modulator with white noise, has the same topological charge and frequency with the Pr field so that they can interfere. The amount of excess noise is controlled by tuning the optical power of the auxiliary field and the white noise added on it. Covariance matrix of the OAM multiplexed CV entangled state is experimentally measured by utilizing two sets of balanced homodyne detectors. In our experiment, the spatially mode-matched local oscillator beams used in the balanced homodyne detectors are obtained from a second set of FWM process in the same vapor cell \cite{PRJ2022}. 

The top rows of Fig. 1(b) and 1(c) show the beam patterns of the Pr and Conj fields for \(l=1\) and \(l=2\), respectively, while the bottom rows show their interference patterns with plane waves at the same frequencies, from which the topological charges of these fields are inferred. It is obvious that the topological charges of the Pr and Conj fields are opposite, which confirms the OAM conservation in the FWM process.

\begin{figure}[t]
	\centering
	\includegraphics[width=10cm]{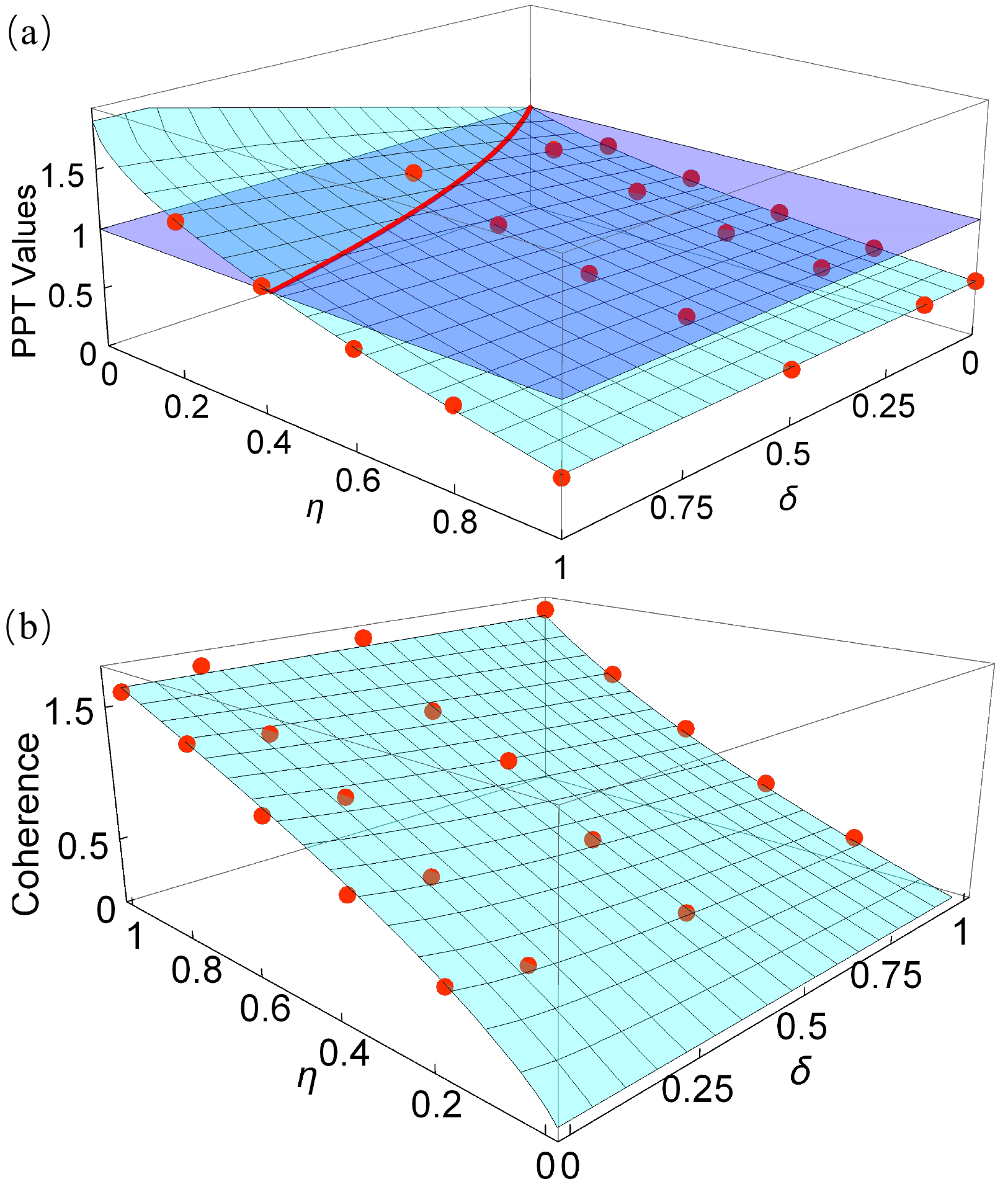}
	\caption{Dependence of PPT values (a) and quantum coherence (b) of the OAM multiplexed CV entanglement on transmission efficiency $\eta$ and excess noise $\delta$ for \(l=2\) in a noisy channel. Curve planes and data points show theoretical predictions and experimental results, respectively. The light blue plane in (a) shows the boundary for entanglement where PPT value is equal to 1, and the red curve shows the boundary where sudden death of entanglement starts to appear.}
\end{figure}

\section{Experimental results and discussion}
In our experiment, the correlation and anti-correlation levels of the initial CV entangled states carrying topological charges \(l=0\), \(l=1\), and \(l=2\) are all around $-3.3$ dB and 6.1 dB, which correspond to $V=0.47$ and $V^{\prime}=4.11$, respectively. The evolution of PPT values and relative entropy of CV bipartite entangled states carrying different topological charges in lossy channels ($\delta=0$) are shown in Fig. 2(a) and 2(b), respectively. The entanglement and quantum coherence between the Pr and Conj fields both degrade as the transmission efficiency decreases. However, the entanglement and quantum coherence are both robust against loss, i.e., they always exist until the transmission efficiency reaches 0. It is obvious that the entanglement and quantum coherence of the CV bipartite entangled states carrying topological charges \(l=1\), \(l=2\) are as robust to loss as their Gaussian counterpart \(l=0\). 

The evolution of PPT values and relative entropy of CV bipartite entangled states carrying different topological charges in a noisy channel with $\delta=1$ (in the units of SNL) are shown in Fig. 2(c) and 2(d), respectively. Compared with the results shown in Fig. 2(a) and 2(b) for the case of lossy channel, both entanglement and quantum coherence degrade faster as the transmission efficiency decreases. Furthermore, entanglement disappears at a certain transmission efficiency of the Pr field $\eta=0.44$ in the presence of excess noise, which demonstrates the sudden death of CV quantum entanglement. In contrast, the quantum coherence always exists until the transmission efficiency decreases to 0. We note that OAM multiplexed CV entangled states carrying high order topological charges \(l=1\), \(l=2\) exhibit the same quantum entanglement and quantum coherence evolution tendency as their Gaussian counterpart \(l=0\) in a noisy channel. 

Figure 3(a) and 3(b) show the dependence of PPT values and relative entropy on transmission efficiency $\eta$ and excess noise $\delta$ for CV bipartite entangled state carrying \(l=2\) in noisy channels. Experimental results at four different amounts of noise  $\delta=0, 0.15, 0.5$ and 1 (in the units of SNL) are taken. It is obvious that both entanglement and quantum coherence degrade with the increase of loss and excess noise. The degradations of entanglement and quantum coherence can be attributed to the incoherent operations of the lossy and noisy channels \cite{Baumgratz2014,Xu2016}. The red curve in Fig. 3(a) shows the boundary where the sudden death of entanglement appears. It is clear that the sudden death of entanglement appears at higher transmission efficiency as the excess noise increases. In contrast, the quantum coherence of the OAM multiplexed CV entangled state still exists even though the entanglement disappears, i.e. quantum coherence of the state is robust against noise. The physical reason for the robustness of quantum coherences in noisy channels is that the proportion of quantum coherence is decreased when it is mixed with thermal noise, but the quantum coherence disappears completely only when infinite thermal noise is involved.

\section{Conclusions}
Here, we experimentally quantify quantum coherence of OAM multiplexed CV entangled state in a noisy channel. Our results demonstrate that quantum coherence of OAM multiplexed CV entangled state is robust against loss and noise, although the sudden death of entanglement is observed at a certain noise level. Recently, it has been shown that entanglement can be transferred in a single-mode cavity \cite{Bougouffa2020}, which is a promising application of robustness of quantum coherence. Our results lay the foundation of applying quantum coherence of OAM multiplexed entangled state in noisy environments.

\begin{backmatter}
\bmsection{Funding}
National Natural Science Foundation of China (NSFC) (No. 11974227, No. 11834010, and No. 61905135); Fundamental Research Program of Shanxi Province (No. 20210302122002); Research Project Supported by Shanxi Scholarship Council of China (2021-003); Fund for Shanxi ``1331 Project" Key Subjects Construction.

\bmsection{Disclosures}
The authors declare no conflicts of interest.

\medskip

\bmsection{Data availability}
Data underlying the results presented in this paper are not publicly available at this time but may
be obtained from the authors upon reasonable request.

\end{backmatter}

\end{document}